\documentclass[acmsmall]{acmart}
\usepackage{amsmath,amsfonts,bm}
\usepackage{graphicx}
\usepackage{booktabs}
\usepackage{pifont}
\usepackage{float}
\usepackage{multirow}
\usepackage{multicol} 
\AtBeginDocument{%
  }

\setcopyright{acmlicensed}
\copyrightyear{2018}
\acmYear{2018}
\acmDOI{XXXXXXX.XXXXXXX}

\acmConference[Conference acronym 'XX]{Make sure to enter the correct
  conference title from your rights confirmation emai}{June 03--05,
  2018}{Woodstock, NY}
\acmISBN{978-1-4503-XXXX-X/18/06}

\begin{document}

\title{Invar-RAG: Invariant LLM-aligned Retrieval for Better Generation}

\author{Ziwei Liu}
\authornote{Both authors contributed equally to this research.}
\email{ziwliu8-c@my.cityu.edu.hk}
\affiliation{%
  \institution{City University of Hong Kong}
  \city{Hong Kong}
  \country{China}
}

\author{Liang Zhang}
\authornotemark[1]
\email{zhangliang@sribd.cn}
\affiliation{%
  \institution{Shenzhen Research Institute of Big Data}
  \city{Shenzhen}
  \country{China}
}

\author{Qian Li}
\email{liqian1@cuhk.edu.cn}
\affiliation{%
  \institution{Shenzhen Research Institute of Big Data}
  \city{Shenzhen}
  \country{China}
}

\author{Jianghua Wu}
\email{wujianghua@sribd.cn}
\affiliation{%
  \institution{Shenzhen Research Institute of Big Data}
  \city{Shenzhen}
  \country{China}
}

\author{Guangxu Zhu}
\email{gxzhu@sribd.cn}
\affiliation{%
  \institution{Shenzhen Research Institute of Big Data}
  \city{Shenzhen}
  \country{China}
}

\begin{abstract}
  Retrieval-augmented generation (RAG) has stood out as a highly promising technique in numerous domains due to its impressive capability of giving reliable answer predictions and addressing the severe hallucination problem. Current RAG systems employ powerful retrieval models to extract external information and leverage large language models (LLMs) as the generation model. Nevertheless, the differences in the architecture between the retriever and generation model have often resulted in notable ineffectiveness, posing a significant obstacle to retrieving and predicting the answers. Recently, LLM-based retrieval has demonstrated its exceptional effectiveness in information retrieval (IR), delivering substantial improvements via LLMs' powerful semantic understanding capability. However, directly applying LLM to RAG systems poses certain challenges. Its massive parametric knowledge impedes the ability to leverage the global information of all the corpus, leading to a scene that current LLM-based retriever usually inputs the summary of documents instead of the whole documents, which causes the feature locality problem. Moreover, various tasks pre-trained in LLMs induce severe variance, which further weakens its performance as the retriever. To address these issues, we propose a novel two-stage fine-tuning architecture called \textbf{Invar-RAG}. In the retrieval stage, by introducing representation learning leveraging the LoRA structure, we construct a LLM-based retriever to address the feature locality problem. Moreover, to consolidate our designed retrieval's performance, we define two patterns (invariant pattern and variant pattern) and propose the invariance loss to eliminate the variance in LLM to some extent. Then, for the generation stage, we devise another fine-tuning method to allow our LLM to better leverage the retrieved information to generate an accurate answer. Experimental results demonstrate that Invar-RAG significantly outperforms existing baselines across three Open-domain Question Answering (ODQA) datasets. Our implementation code is available in \textbf{Supplementary Material} to ease reproducibility.
\end{abstract}


\received{20 February 2007}
\received[revised]{12 March 2009}
\received[accepted]{5 June 2009}

\maketitle

\section{Introduction}
    Over the past decade, large language models (LLMs) have demonstrated promising capability in processing natural language~\cite{3}. Owing to the vast amount of knowledge encoded in their internal parameters, LLMs such as GPT~\cite{1} and LLaMa~\cite{2} have demonstrated remarkable performance on various downstream tasks, including Open-domain Question Answering (ODQA)~\cite{4}, Reading Comprehension~\cite{5}. However, the fixed parametric knowledge of LLMs has hindered the further applications of LLMs and made them prone to errors (hallucination~\cite{6} and factual errors~\cite{7}).
    To overcome the limitations of parametric knowledge, one promising approach is Retrieval-Augmented Generation (RAG)~\cite{7,8}. Compared to relying solely on parametric knowledge, RAG enables LLMs to use retrievers to access relevant information from external knowledge sources, enhancing their question-answering abilities. Among the two components of RAG, current methods primarily focus on optimizing the retriever to return more relevant documents due to the high cost of fine-tuning and black-box LLM APIs. Previous retrievers leveraged deep learning technology (e.g., dense retrieval~\cite{9}) to encode the text representations from the lexical space into the high-dimensional latent space, allowing them to model more complex semantic relationships between queries and corpora. However, the separation between the retriever and generation model has hindered their full integration, limiting their compatibility in downstream applications~\cite{}. Some advanced RAG systems, such as RA-DIT~\cite{10}, have adopted joint training mechanisms that fine-tune both the retriever and the generation model for better alignment. However, this approach is impractical due to the need for frequent fine-tuning and fails to utilize the LLMs' semantic understanding capabilities during the retrieval stage.
    Consequently, generative retrieval (GR), also known as LLM-based retrieval, leverages the parametric memory of generative models to directly generate document identifiers (DocIDs)~\cite{15}, which has aroused much attention. By memorizing the documents as the parametric knowledge of LLM, this kind of method breaks the limitations of traditional IR in terms of document granularity and simple relevance matching~\cite{38}, offering more flexibility and creativity, thus better meeting practical needs.
However, two severe problems hinder the current LLM-based retrieval. 1)\textbf{Feature Locality}: LLM-based retrieval normally adopt language models to learn the mapping from queries to the relevant document DocIDs. However, these DocIDs actually can not fully represent the global information of the passages. Meanwhile, directly feeding the whole passages into LLM is costly and infeasible, causing a trade-off between effectiveness and efficiency. 2)\textbf{Retrieval Variance}: Due to the inherent generative inconsistency property of large language models~\cite{7}, current LLM-based retrieval may generate unforeseeable variances, especially when the input query or the size of context varies, directly results in undesirable and vulnerable performance which may not be preferred.

\begin{figure*}[!t]
        \label{fig:Overview}
	\centering
 
	\includegraphics[width = \linewidth]{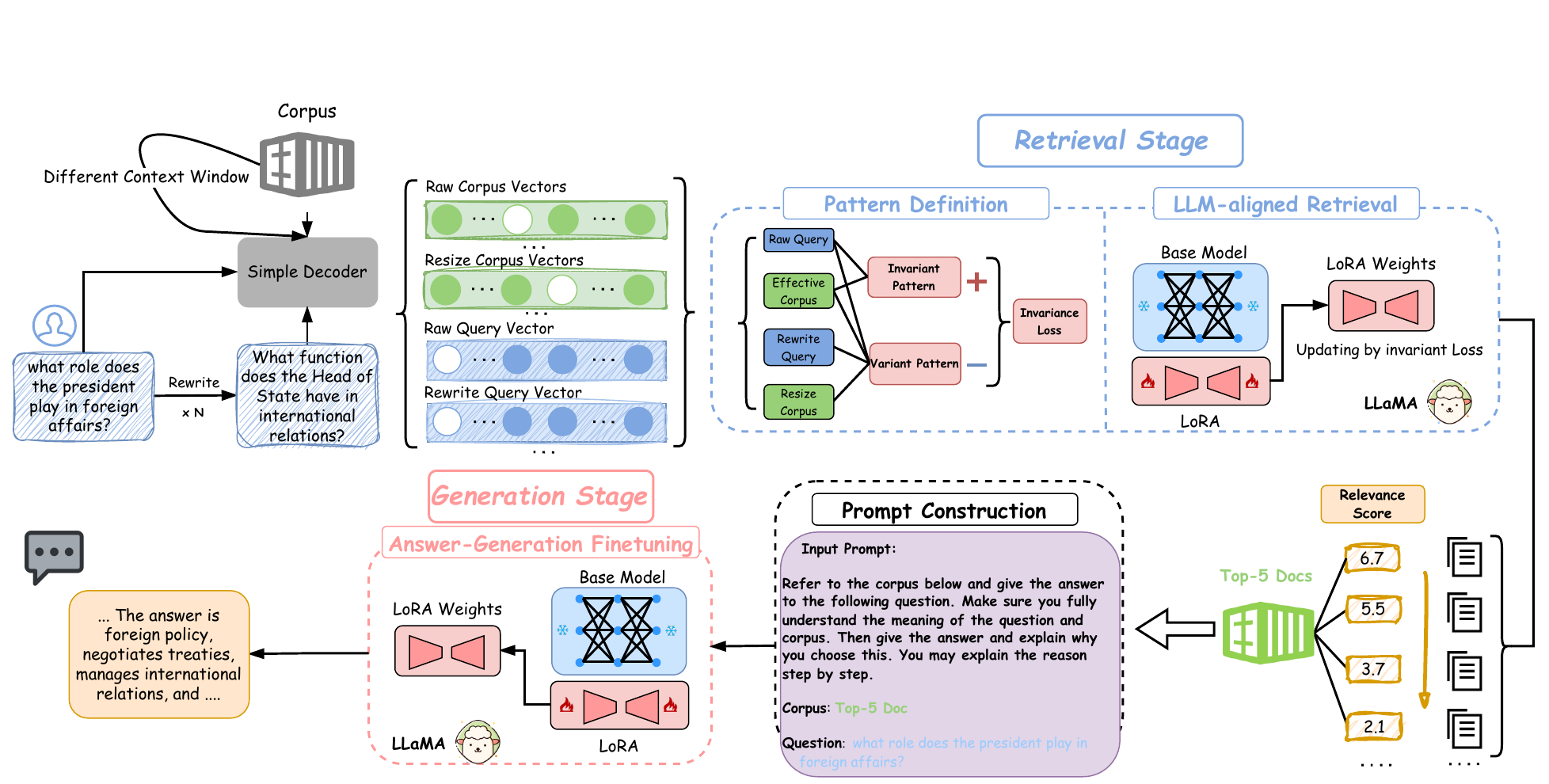}
	\caption{Overview of proposed Invar-RAG.}
      
\end{figure*}

Considering the problems mentioned above and better leveraging the capability of LLMs, we propose a fully LLM-based architecture with a two-stage fine-tuning method called Invar-RAG, as illustrated in Figure~\ref{fig:Overview}. In the retrieval stage, our approach initializes the pre-trained LLaMA~\cite{2} as the backbone and follows the bi-encoder architecture in DPR~\cite{11} to construct our retriever. Compared to normal GR methods which need an iterative process of encoding and decoding, we introduce a component called LLM-aligned Retrieval. It first represents the input query and corpora into high-dimension space using a small language model (MiniLM)~\cite{12}, then introduces a new loss function constructed by KL-divergence to align the coarse query-documents pairs representation to the LLM's representation space. This allows the retriever to leverage the rich prior knowledge of LLM, typically addressing the feature locality caused by only feeding DocIDs to LLM. Moreover, based on the initial objective of our LLM-aligned Retrieval, we introduce the invariance loss to overcome the variance in the retrieval stage. By recognizing the invariant pattern that contributes the most to the performance and gradually forcing the model to rely on the invariant pattern, we can avoid the unforeseeable variances in practice and enhance the robustness of our RAG system. Finally, in the generation stage, we freeze the weights we fine-tuned before and optimize the generation function to allow the LLM to give correct answers to the retrieved documents. Our contributions are summarized as follows: \begin{itemize}
    \item We introduce Invar-RAG, a novel framework featuring a two-stage fine-tuning method on a single shared LLM, including the retrieval stage and generation stage.
    \item We introduce a novel LLM-based retrieval method containing representation learning and invariance loss, respectively addressing the issues of feature locality and retrieval variance.
    \item We validate Invar-RAG's performance on three public ODQA datasets, no matter for retrieval performance or generation performance, demonstrating its superiority.
\end{itemize}

\section{Methodology}
In this section, we introduce a novel retrieval-augmented language model architecture, Invar-RAG, which addresses the previously mentioned issues by using LLM-aligned retrieval combined with a specially designed invariance loss. We first present an overview of our proposed architecture, followed by a detailed explanation of its key components, and finally, we introduce how we construct the prompts.
\subsection{Overall Framework of Invar-RAG}
In this section, we provide an overview of Invar-RAG, as shown in Fig.~\ref{fig:Overview}. We begin by using query rewriting and context window resizing to introduce various types of variance. Next, we apply a small language model to map these texts into a high-dimensional vector space, generating coarse representations. We then adopt LLM-aligned retrieval to align the coarse representation with the LLM's representation and compute the basic relevance score via dot product. This approach addresses the feature locality problem by feeding the entire document representation into the LLM, rather than a single DocID. Additionally, to address feature variance, we define an invariance loss based on the initial KL-divergence loss function in representation learning, encouraging the model to rely on invariant patterns. Finally, by constructing appropriate prompts and fine-tuning the generation stage, we optimize the LLM to better utilize the retrieved information, generating more accurate answers to the given questions.\\
\subsection{Retrieval Stage}
\label{sec:re}
\noindent \textbf{\textit{Architecture.}}
For the Retriever architecture, we follow the approach of previous work~\cite{13}, using the bi-encoder architecture from DPR~\cite{11}, but replacing the backbone model with LLaMA~\cite{2}. 
Considering the efficiency, we first compute the vector embedding of a document ${d}^R_{i} \in \mathcal{D}^R$ as:
\begin{equation}
    {V}_{r} ({{d}^R_{i}})= \operatorname{Decoder}\left(`t_1␣t_2␣\cdots␣t_k'\right)\left[-1\right]
\end{equation}
Where k represents the maximum number of trunks, and $\operatorname{Decoder}(\cdot)$ represents the embedding layer of a small language model (MiniLM-v2), which maps the trunks $(t_1␣t_2␣\cdots␣t_k)$ from the initial text space to a high-dimensional dense vector space. For the vector embedding of query $q$, we leverage our LLM structure to return the last layer token representation as the representation, denoted as $V_q$.

To leverage the LLM's prior knowledge while maintaining efficiency, we further align the query-document pairs to the LLM's representation space and denote the processed document representation as $V_p({{d}_i^R})$. Consequently, we can compute in terms of the dot product to get the two relevance scores:
\begin{equation}
\begin{aligned}
  s_{raw}\left(V_q,V_r({{d}_i^R})\right) &= V_q \cdot V_r({{d}_i^R}) \\
  s_{pro}\left(V_q,V_p({{d}_i^R})\right) &= V_q \cdot V_p({{d}_i^R})
\end{aligned}
\end{equation}
where the basic relevance score between the query and documents processed by small LM denotes as $P_{raw}$ and the target relevance score computing between query and LLM-processed documents denotes as $P_{pro}$.

\noindent \textbf{\textit{LLM-aligned Retrieval.}}
Building on the above structure, we enhance our LLM-based retriever's ability to return more relevant documents. We introduce LLM-aligned retrieval with invariance loss in the retrieval stage, effectively addressing the aforementioned issues.
Current alignment methods, such as RA-DIT~\cite{10}, primarily focus on aligning the scoring functions between the retriever and the generator. However, the initial structural differences between novel retrievers (e.g., DRAGON+~\cite{14}) and LLMs still impede further optimization of the overall RAG system. Therefore, we design a novel LLM-based retriever to resolve this issue.
Unlike previous LLM-based GR methods~\cite{15}, we no longer need to use DocIDs to retrieve relevant documents, which may cause the feature locality issue mentioned earlier. Instead, we adopt a fine-tuned alignment process that enables the LLM to perform representation learning. 
We leverage LoRA architecture~\cite{16} to add additional adapter parameter $\theta_\mathcal{R}$ to our raw representation $V_r$, denoted as $V_r({{d}_i^R},\theta_\mathcal{R})$. 
The corresponding relevance score can then be re-normalized among top-k relevant chunks $\mathcal{D}^{{R}'} \subset \mathcal{D}^{{R}}$ as:
\begin{equation}
\label{equ:rs}
    S^r_{\mathcal{R}}(V_r({{d}_i^R},\theta_\mathcal{R})|V_q) = \frac{\exp {s_{raw}\left(V_q,V_r({{d}_i^R},\theta_\mathcal{R})\right)}}{\sum_{{d_i^{{R}}}'\in \mathcal{D}^{{R}'}} \exp {s_{raw}\left(V_q,V_r({{{d}_i^R}'},\theta_\mathcal{R})\right)} }
\end{equation}
For each document in the corpus, we need to compute $S^r_{\mathcal{R}}$ for $n$ times ($n$ represents the number of documents in $\mathcal{D}^R$) to rank the relevance scores. Consequently, the initial loss function for representation learning can then be defined by minimizing the KL-divergence~\cite{17} of two relevance scores leveraging Eq.~\ref{equ:rs}.
\begin{equation}
    \label{equ:kl}
    \mathcal{L}_{rl}\left(\mathcal{D}^R\right ) = \mathbb{E}_{d_i^R \in \mathcal{D}^R}\operatorname{KL}(S^r_{\mathcal{R}}(V_r({{d}_i^R},\theta_\mathcal{R})|V_q) \| S^r_{\mathcal{R}}(V_p({{d}_i^R})|V_q))
\end{equation}
Following the previous works~\cite{10,19}, fine-tuning both encoders hurt the performance~\cite{18}, we only update a part of our initialized retriever, which is in charge of computing the query representation.\\
\noindent \textbf{\textit{Invariance Loss.}}
To further enhance retrieval accuracy while maintaining robustness, we introduce invariance loss, building on our initial KL-divergence loss. Current refinement methods primarily rely on query rewriting~\cite{20,21} or LLM generation~\cite{22} to expand the search space and re-rank document chunks. However, they fail to recognize the effectiveness of different rewriting and generation procedures, directly resulting in the invariance problem. Specifically, we begin by rewriting the query and adjusting the context window to broaden the search space. Since not all refinement methods are effective, we identify invariant patterns to preserve retrieval performance while gradually incorporating weighted variant patterns to broaden the search space. To achieve this, we use the LSR score from LM-Supervised Retrieval~\cite{23} to determine whether a document effectively enhances the LLM's answer prediction capability.
For a training sample $(q,y)$, where q and y respectively represent the input query and output result, we first define the output probability of LM as:
\begin{equation}
\label{equ:Plm}
    p_{LM}(y|{V}(d_i^R \circ x))= \sum_{{d}^R_{i} \in \mathcal{D}^R} p_{LM}(y|{V}({d}^R_{i} \circ x))\cdot P_{\mathcal{R}}({d}^R_{i}|x) 
\end{equation}
Then, for the LSR score for a retrieved document $d_i^R$:
\begin{equation}
    \label{equ:LSR}
    P_{LSR}(d_i|q,y)= \frac{\exp( p_{LM}(y|d_i \circ q)/\tau)}{\sum_{d_i'\in {\mathcal{D}^R}'}\exp( p_{LM}(y|d_i' \circ q)/\tau)} \approx \frac{\exp( p_{LM}(y|d_i \circ q)/\tau)}{\sum_{d_i'\in \mathcal{D}^R}\exp( p_{LM}(y|d_i' \circ q)/\tau)}
\end{equation}
where $\tau$ is the temperature hyperparameter of LLM, $\mathcal{D}_R' \subset \mathcal{D}_R$ denotes the top-k retrieved trunks. 
Assuming the query after rewriting as $q^r$, documents set after resizing as $\mathcal{D}_R^{re} =({d}^{re}_1,{d}^{re}_2,\cdots,{d}^{re}_n)$, we can leverage the Eq.~\ref{equ:LSR} to calculate the score matching from (1) $q$ to $d_i$, (2) $q^r$ to $d_i$, (3) $q$ to ${d}^{re}_i$ and (4) $q^r$ to ${d}^{re}_i$. We recognize the invariant pattern as the top-$l$  ranked documents, denoted as $\mathcal{D}_{in}$, where $0< l < k$ is satisfied. For other documents, we assume them as variant pattern $\mathcal{D}_{var}$, which contribute little to generating effective answers. 

The invariant loss function can be formalized as follows:
\begin{equation}
\label{equ:invar}
\mathcal{L}_{invar}(\mathcal{D}_{in}) = \operatorname{Var}_{\mathcal{D}\subseteq\mathcal{D}_{var}}(\mathbb{E}_{d_{in}^R \in (\mathcal{D}_{in} \cup \mathcal{D})} \operatorname{KL}(P^r_{\mathcal{R}}(V_r(d_{in}^R,\theta_\mathcal{R})|V_q) \|P^p_{\mathcal{R}}(V_p(d_{in}^R)|V_q)))
\end{equation}
This invariance loss measures the variance of the model’s aligning ability under multiple interventions (i.e., query rewriting and context resizing) by only allowing the documents in the $\mathcal{D}_{invar}$ to update the loss. The whole training objective can then be presented as:
\begin{equation}
\label{equ:loss}
\min_{\theta_R} \mathcal{L}_{rl} + \lambda \mathcal{L}_{invar}
\end{equation}
where the task loss $\mathcal{L}_{rl}$ is minimized to align the two different representations while the $\mathcal{L}_{invar}$ enables the model to rely more on the invariant pattern, and $\lambda$ is a hyperparameter to balance between two objectives.

\subsection{Generation Stage}
\label{sec:gene}
In the generation stage, We followed the same architecture as in the retrieval stage for answer prediction.
To improve the generative capability of LLM for leveraging the retrieved information better, followed by prior works~\cite{10,23}, we adopt another LoRA adapter to fine-tune our model on different tasks. Specifically, for the same training sample $(x,y)$, we retrieve the \textit{top-}$\overline{k}$ relevant document chunks $\mathcal{D}_G' \subset \mathcal{D}_G$ by performing our model on retrieval task. For each retrieved chunk $d_i \in \mathcal{D}_G'$, we design a special fine-tuning example by prepending it to the prompt as background information and create k independent instances for one original example: $\left\{ (d_i \circ x, y)|i=1,\cdots,\overline{k} \right\}$. Then, following the previous work~\cite{10,24}, we fine-tune the language model using the next-token prediction objective and minimize the loss as follows:
\begin{equation}
    \mathcal{L}(\mathcal{D}_G')= -\sum_i \log p_{LM}(y|d_i \circ x)
\end{equation}
By applying this fine-tuning method, the generation stage benefits in two ways: (i) it improves the model's performance on the generation task by providing more accurate predictions based on the retrieved information; (ii) when the retrieved documents fail to provide an accurate answer, the approach enables the LLM to rely on its parametric knowledge to generate an answer while disregarding misleading retrieved documents.
\section{Experiment}
In this section, we will first introduce the experiment setting. Then we present extensive experiments to evaluate the effectiveness of our proposed Invar-RAG architecture in different stages (retrieval and generation). All the reported experimental results are the average values obtained from five independent runs of the algorithm.
\subsection{Setting}
\subsubsection{Datasets}
Following the prior works~\cite{10,25}, we choose two ODQA datasets (FreebaseQA~\cite{26} and MS-MARCO~\cite{27}) and one reading comprehension (RC) dataset to do the representation learning in the retrieval stage (denoted as $\mathcal{D}_R$) while leveraging other three datasets (Web Question~\cite{28}, Wiki Question Answering~\cite{29}) and SQuAD v2\footnote{\url{https://rajpurkar.github.io/SQuAD-explorer/explore/v2.0/dev/Prime_number.html}} for fine-tuning the LLM in the generation stage (denoted as $\mathcal{D}_G$). The statistic of chosen datasets is shown in Tab.\ref{tab:dataset}. For detailed descriptions and complied templates, please refer to \textbf{Appendix A}.
\begin{table*}[t]
\vspace{-4mm}
\centering
\caption{The Statistics of Fine-tuning Datasets.}
\resizebox{\linewidth}{!}{
\begin{tabular}{cccccc}
\toprule[1pt]
Dataset &  HF identifier & $\mathcal{D}_{R}$ & $\mathcal{D}_{G}$ &Training Sample&Task\\ 
\midrule
Wiki QA~\cite{29} & {wiki\_qa} & \ding{55} & \ding{51} &20360& Open-domain QA \\
FreebaseQA~\cite{26} & {freebase\_qa} & \ding{51} & \ding{55} & 20358&Open-domain QA  \\
MS-MARCO~\cite{27} & {ms\_marco} & \ding{51} & \ding{55}& 80143&Open-domain QA \\
Web Question~\cite{28} & {web\_question} & \ding{55} & \ding{51} &3778& Open-domain QA \\

\midrule
SQuAD v2 & {squad\_v2} & \ding{55} & \ding{51}& 130319&Reading Comprehension \\
\bottomrule[1pt]
\end{tabular}
}
\label{tab:dataset}
\vspace{-4mm}
\end{table*}

\subsubsection{Evaluation}
To access our performance, we conduct the evaluation on four knowledge-intensive datasets, such as TriviaQA (denoted as TQA)\footnote{\url{https://nlp.cs.washington.edu/triviaqa}}, Natural Question (denoted as NQ)\footnote{\url{https://ai.google.com/research/NaturalQuestions}} and PopQA\footnote{\url{https://huggingface.co/datasets/akariasai/PopQA}}, that are not involved in the training progress. For the evaluation metric, we evaluate our model's generation performance using the Exact Match~\cite{30}, which indicates whether gold answers are included in the model generations followed by the setting in prior work~\cite{10,31}. Furthermore, to evaluate our proposed retriever's performance, we employ the Acc@5 and Acc@20 as evaluation metrics, which are widely used in related studies~\cite {32,33}. These metrics assess the proportion of questions where the correct answers appear in the
top-5 or top-20 retrieval results, offering a comprehensive evaluation of the retrieval performance. For more details, please refer to the description and methods in \textbf{Appdendix B}.

\subsubsection{Implementation Details}
In this section, we provide a detailed description of our framework's implementation. The code can be found in \textbf{Supplementary Material}. 
For both the retrieval and generation stages, the LLaMA-2-7B checkpoint\footnote{\url{https://huggingface.co/meta-llama/Llama-2-7b-hf}} is leveraged to initialize the pre-trained weights of our architecture. For the GPU selection, We perform our further fine-tuning on $4 \times 40$G NVIDIA V100 GPUs.

\textbf{Retrieval Stage.}
Following the previous work's setting~\cite{13}, as LLaMA is a decoder-only architecture, we append an end-of-sequence token \textless EOS\textgreater ~to the input sequence and regard the last layer representation as the dense representation to calculate the similarity score. Considering the possible effect caused by the size of each dense representation, we also employ the normalization procedure to map the original representation into unit vectors during both the training and inference stages.
For the fine-tuning progress in the retrieval stage, we adopt LoRA architecture~\cite{16} to reduce the high cost of GPU memory. The detailed hyperparameters we used can be found in \textbf{Appendix C}.

\textbf{Generation Stage.}
We hold $\overline{k}$ in Sec.\ref{sec:gene} equal to 5 to generate instances for a single example and append multiple examples together to improve the efficiency (the length is limited to 4096 tokens). The used hyperparameters are also shown in the \textbf{Appendix C}. Other implementation details are the same as original papers~\cite{10,23}.
\subsubsection{Baselines}
To demonstrate the effectiveness of our proposed architecture, we compare the retrieval performance of our Invar-RAG with state-of-the-art retrieval methods, including sparse retrieval (BM25~\cite{34}), dense retrieval (BGE~\cite{35}, Contriever~\cite{33}) and LLM-based retrieval (LLM-embedder~\cite{36} and RepLLaMA~\cite{13}). Furthermore, for the corresponding RAG performance, we conduct extensive experiments compared to the novel retriever + generation model to show our superiority. The descriptions for each baseline are listed in the \textbf{Appendix D}.
\subsection{Overall Performance}
In this section, we present performance comparison experiments on two stages, respectively, with three knowledge-intensive ODQA datasets. The results show that our Invar-RAG architecture outperforms all competing sparse, dense, and LLM-based baselines in retrieval and their downstream RAG in generation. Such a comparison highlights the effectiveness of our unique design for two-stage fine-tuning.
\begin{table*}[]
\centering
    \caption{Retrieval performance comparison between our designed retriever in Invar-RAG and other baselines. The best results are bold, and the second-best are underlined.}
    \label{table:retrieval}
    \resizebox{\textwidth}{!}{%
    \footnotesize 
\begin{tabular}{lcccccc}
\toprule[1pt]
\multicolumn{1}{c}{\multirow{2}{*}{Models}} & \multicolumn{2}{c}{TQA}       & \multicolumn{2}{c}{NQ}        & \multicolumn{2}{c}{PopQA}     \\ \cline{2-7} 
\multicolumn{1}{c}{} &
  \multicolumn{1}{l}{Acc@5} &
  \multicolumn{1}{l}{Acc@20} &
  \multicolumn{1}{l}{Acc@5} &
  \multicolumn{1}{l}{Acc@20} &
  \multicolumn{1}{l}{Acc@5} &
  \multicolumn{1}{l}{Acc@20} \\ \midrule
BM25~\cite{34}                                        & 62.5          & 73.0          & 49.0          & 67.0          & 35.5          & 51.5          \\
BM25+BGE(re-ranker)~\cite{32}                                 & \underline{72.5}    & 78.0          & 68.0          & 76.5          & 54.0          & 60.0          \\
Contriever~\cite{33}                                  & 68.0          & \underline{80.5}    & 68.0          & 84.0          & 62.0          & 77.5          \\
BGE-base~\cite{35}                                   & 69.5          & 80.0          & \underline{77.0}    & 86.0          & \underline{72.0}    & \textbf{83.0} \\

LLM-embedder~\cite{36}                                & 67.5          & 77.5          & 75.5          & \underline{86.5}    & 70.0          & 79.5          \\
RepLLaMA~\cite{13}                                    & 66.5          & 76.0          & 72.0          & 85.5          & 68.5          & 74.5          \\
\textbf{Invar-retrieval (ours)}                      & \textbf{74.0} & \textbf{81.5} & \textbf{80.5} & \textbf{88.0} & \textbf{73.5} & \underline{82.5}    \\ \midrule
Improv.                                     & 2.1\%        & 1.2\%        & 4.6\%        & 1.7\%        & 2.1\%        & -0.6\%        \\ \bottomrule[1pt]
\end{tabular}}
\vspace{-4mm}
\end{table*}
\subsubsection{Retrieval Performance}
\label{sec:re_p}
In this section, we will present and analyze the retrieval performance of our designed architecture. As illustrated in Tab.\ref{table:retrieval}, the sparse retriever BM25 fails to map the given text to proper representations. Although employing an additional model as the re-ranker improves the performance to some extent, the retrieval capability remains sub-optimal due to the inferiority of BM25.  Besides, Novel dense retrievers, like BGE and Contriever, present comparable performance over the three datasets, suggesting their effectiveness in leveraging contrastive learning or task-specific fine-tuning. However, they still slightly lag behind our designed Invar-retrieval because of the neglect of rich semantic information~\cite{13}. Current researchers have proposed several LLM-based retrievers i.e., LLM-embedder~\cite{36} and RepLLaMA~\cite{13}, which leverage the rich prior knowledge that LLM initially has. However, due to the high cost of processing the massive corpus, it is infeasible to handle all the chunks within the LLM. Moreover, the variance problem that happens in LLM also leads to relatively inferior performance. Correspondingly, we propose our LLM-based retrieval model, Invar-retrieval, as a part of our designed Invar-RAG. The results shows that our methods outperform all the sparse, dense and LLM-based retrievers, especially under the Acc@5 measurement, contributing to our designed invariance loss in reducing the variant and ineffective patterns.

\subsubsection{Generation Performance}
In this section, we will analyze the answer generation capability for our designed Invar-RAG. Based on the astonishing performance of our designed Invar-retrieval, we further fine-tune the language model to leverage the retrieved documents for better question-answering capability. From the experimental results presented in Tab.\ref{table:generation}, our Invar-RAG shows reasonable performance on the three ODQA datasets, echoing the performance of the retrievers designed in Tab.\ref{table:retrieval}.
\begin{table*}[]
\centering
    \caption{Generation performance comparison between our designed Invar-RAG and other baselines. The best results are bold, and the second-best are underlined.}
    \label{table:generation}
    
\begin{tabular}{lccc}
\toprule[1pt]
\multicolumn{1}{c}{\multirow{2}{*}{Models}} & TQA & \multicolumn{1}{l}{PopQA} & \multicolumn{1}{l}{NQ} \\ \cline{2-4} 
\multicolumn{1}{c}{}             & \multicolumn{3}{c}{Exact Match} \\ \midrule
BGE-base + LLaMA-2-7B            & \underline{74.1}     & 49.8     & 52.1     \\
BM25 + BGE(re-rank) + LLaMA-2-7B & 72.3    & 48.2     & 51.6     \\
LLM-embeder + LLaMA-2-7B         & 71.8    &  \underline{51.1}    & \underline{54.1}     \\
Contriever + LLaMA-2-7B          & 72.6     & 48.6     & 51.8      \\
Invar-RAG                        & \textbf{75.3}     & \textbf{53.6}     & \textbf{56.2}    \\ \midrule
Improv.                          & 1.6\%      & 4.9\%      & 3.9\%      \\ \bottomrule[1pt]
\end{tabular}
\vspace{-4mm}
\end{table*}

\subsection{Ablation Study}
In this section, we analyze the efficacy of the two-stage fine-tuning in the Invar-RAG architecture, including the retrieval stage (\textit{LLM-aligned Retrieval with Invariance Loss}) and the generation stage. We design three variants: (1)\textit{w/o representation learning}: this variant uses the coarse text representation mapped by small language model (MiniLM-v2~\footnote{\url{https://huggingface.co/sentence-transformers/all-MiniLM-L6-v2}}) to calculate the relevance score and adopt the same generation fine-tuning method in Sec.\ref{sec:gene}. (2)\textit{w/o invariance loss}: the second variant leverages the KL-divergence loss without the additional invariance loss to perform the representation learning. (3)\textit{w/o generative fine-tuning}: this variant directly feeds retrieved documents and the corresponding question as a prompt to generate the answer. The fine-tuning datasets for each variant we used are presented in Tab.\ref{tab:abla_dataset}.
From the performance comparison in Tab.\ref{tab:abs}, We can conclude that: \begin{itemize}
    \item With the representation learning method, LLM-based retrieval contributes to improving the retrieval and corresponding generation performance. 
    \item  Invariance loss significantly boosts our designed Invar-RAG by making the prediction rely more on invariant patterns. 
    \item Generative fine-tuning is crucial for enhancing LLM's capability of giving predictions based on retrieved information. Moreover, it shows the effectiveness of the two-stage fine-tuning for a single LLM.
\end{itemize}
\begin{table*}[]
\centering
\vspace{-8mm}
\caption{The Statistics of Datasets in Ablation study.}

\resizebox{\linewidth}{!}{
\begin{tabular}{lccccc}
\toprule[1pt]
\multicolumn{1}{c}{\multirow{2}{*}{Model Variants}} &
  \multicolumn{2}{c|}{Retrieval Fine-tuning} &
  \multicolumn{3}{c}{Generation Fine-tuning} \\ \cline{2-6} 
\multicolumn{1}{c}{} &
  \multicolumn{1}{l}{Freebase QA} &
  \multicolumn{1}{l|}{MS-MARCO} &
  \multicolumn{1}{l}{Wiki QA} &
  \multicolumn{1}{l}{Web Question} &
  \multicolumn{1}{l}{SQuAD v2} \\ \midrule
Default & \ding{51} & \ding{51} & \ding{51} &\ding{51}  &\ding{51}  \\
w/o representation learning & \ding{55} & \ding{55} & \ding{51} & \ding{51} & \ding{51} \\
w/o invariance loss & \ding{51} & \ding{51} & \ding{51} & \ding{51} & \ding{51} \\
w/o generative fine-tuning  & \ding{51} & \ding{51} & \ding{55} & \ding{55} & \ding{55} \\ \bottomrule[1pt]
\end{tabular}
}
\label{tab:abla_dataset}
\end{table*}
\begin{table}[]
\centering
\vspace{-4mm}
\caption{Ablation Study on TQA.}
\begin{tabular}{lccc}
\toprule[1pt]
\multicolumn{1}{c}{\multirow{2}{*}{Model Variants}} & \multicolumn{2}{c|}{Retrieval} & Generation \\
\multicolumn{1}{c}{}                                & Acc@5     & \multicolumn{1}{c|}{Acc@20}    & Exact Match            \\ \midrule
Default                     & 74.0 & 81.5 & 75.3 \\
w/o representation learning & 63.5 & 73.6 & 74.1 \\
w/o invariance loss         & 71.5 & 81.0 & 74.6 \\
w/o generative fine-tuning  &  /&/  & 73.4 \\ \bottomrule[1pt]
\end{tabular}
\vspace{-4mm}
\label{tab:abs}
\end{table}
For ablation results on other two datasets (NQ and PopQA), please found them in \textbf{Appendix E}.
\subsection{Invariance Analysis}
In this section, we leverage a special example in TQA to illustrate the effectiveness of our designed invariance loss in two parts: (i) the importance of defining different patterns, (ii) the difference in retrieval performance that invariance loss brings. 

As mentioned in Sec.~\ref{sec:re}, we return four different sets of retrieved documents and rerank them by LSR score to identify the invariant pattern. There are two reasons to explain this: \begin{itemize}
    \item Rewriting the query and resizing the context window does affect the normal relevance score computing by the dot product, leading to the variance in practice when we feed different lengths or formats of questions to the RAG system to ask for the answer. 
    \item Prior works~\cite{37} have shown that adding a suitable amount of irrelevant or relatively ineffective documents does help improve the retrieval performance.
\end{itemize}
\begin{figure*}[!t]
        \label{fig:pattern}
	\centering
	\includegraphics[width = \linewidth]{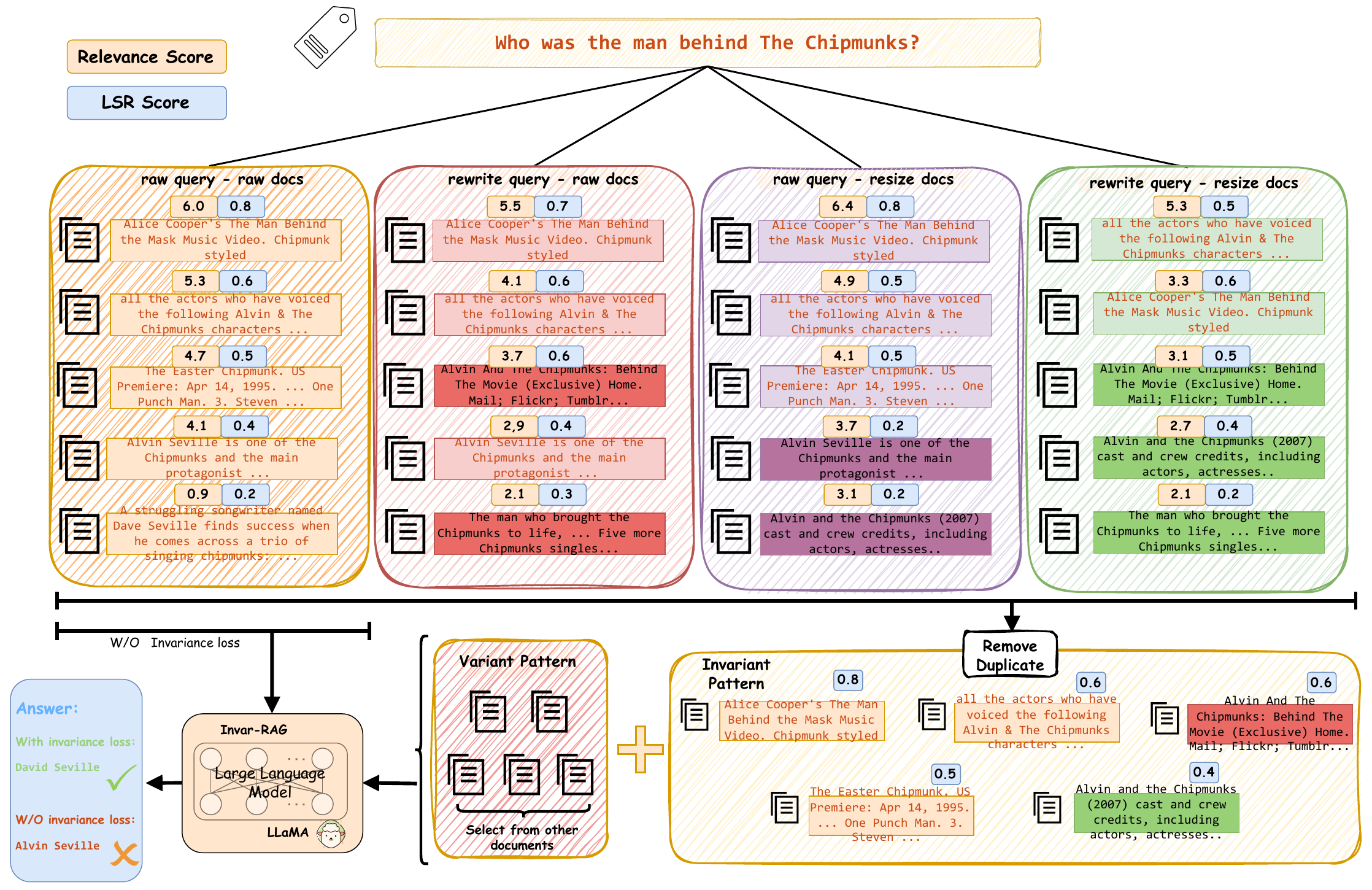}
	\caption{Special example for illustrating the effectiveness of invariance loss.}
      \vspace{-4mm}
\end{figure*}
To verify that, we present the normal relevance score and LSR score of each retrieved document in four different sets in Fig.~\ref{fig:pattern}. The darker color represents the change that happened in the Top-5 documents. We can see that, for the question: `Who was the man behind The Chipmunks', the relevance score for the top-5 documents in each set shows substantial changes while the LSR score does not vary a lot, which means the variance caused by rewriting query or resizing context window change the importance of documents, directly resulting in poor retrieval performance.
Moreover, as illustrated in Fig.~\ref{fig:pattern}, the variant without invariance loss gives the wrong answer to our selected question example while our designed Invar-RAG system accurately predicts the result.
\section{Related Work}
\textbf{Information Retrieval:} Advancements in deep learning have revolutionized information retrieval systems, enhancing their personalization and accuracy in retrieving relevant documents. Early information retrieval frameworks employed sparse retrievers~\cite{34} or dense retrievers~\cite{33,35} to represent large corpora but struggled to capture deep semantic relationships~\cite{11}. LLM-based retrievers (generative retrieval) have since emerged as notable methods, leveraging the rich prior knowledge of LLMs to significantly improve performance by converting documents into parametric knowledge and generating them instead of computing similarity scores~\cite{4}. However, the frequent encoding and decoding processes in LLMs severely hinder efficiency~\cite{4}. To address the trade-off between effectiveness and efficiency, we propose invar-retrieval in our architecture, enabling the model to efficiently retrieve the most relevant documents without introducing variance.

\textbf{Retrieval-augmented Language Model:} Currently, retrieval-augmented language models have proven effective in answering questions by leveraging external information through the integration of novel retrievers and LLMs~\cite{4}. However, the architectural gap between retrieval and generation continues to hinder unified optimization across the entire retrieval-augmented generation system~\cite{7}. To address the isolation between retrieval and generation, a novel architecture called RA-DIT was introduced~\cite{10}. By aligning retriever scoring with LSR scoring~\cite{23}, it has been shown to deliver state-of-the-art performance across various tasks. However, it still employs dense retrievers like DRAGON+~\cite{14} in the retrieval stage, which fails to eliminate the problem at its source and introduces inefficiencies throughout the process. Correspondingly, we introduce a representation learning method and invariance loss in our Invar-RAG architecture, which partially addresses these issues and explores a novel approach to using a single LLM for multiple roles within the RAG system.
\section{Conclusion}
In this paper, we analyze the challenges and problems of current methods to apply the large language model as a retriever in the  RAG system and propose a novel framework, Invar-RAG, to address these challenges. We introduce an LLM-aligned retrieval method, incorporating a well-designed representation learning approach to align coarse query-document pairs with the LLM's representation space, allowing our architecture to leverage the extensive parametric knowledge of the LLM to compute relevance scores. Additionally, to address retrieval variance, we propose invariance loss, building on our initial KL-divergence loss, during the retrieval stage to reduce the impact of irrelevant documents. Finally, we perform additional fine-tuning on the same LLM for the answer-generation task, enabling our architecture to better utilize the retrieved information and provide more accurate predictions. Extensive experiments on three open-domain question-answering datasets confirm Invar-RAG's superiority and validate the effectiveness of each module.

\bibliographystyle{ACM-Reference-Format}

\appendix
\section{Appendix A: Datasets Information}
For training and evaluating our RAG system, we select eight datasets (five for training and three for evaluation). All the datasets we used are downloaded from the Hugging Face website\footnote{\url{https://huggingface.co/datasets}}.
\subsection{Fine-tuning Datasets}
We choose five datasets to fine-tune our two-stage architecture. Specifically, in the retrieval stage, we choose two Open-domain Question Answering (ODQA) datasets, i.e., FreebaseQA~\cite{2} and MS-MARCO~\cite{3}, to fine-tune our model for returning more relevant documents leveraging the powerful semantic understanding capability of LLM while avoiding the variance that happened in the practical situation.
\begin{itemize}
    \item \textbf{FreebaseQA}\footnote{\url{https://huggingface.co/datasets/microsoft/freebase_qa}}: This dataset is designed for open-domain factoid question answering using the Freebase knowledge graph. It contains 28,348 trivia-style questions and over 54,000 question-answer matches from TriviaQA and trivia website.
    \item \textbf{MS-MARCO}\footnote{\url{https://huggingface.co/datasets/microsoft/ms_marco}}: The MS-MARCO (Microsoft MAchine Reading Comprehension) is a large-scale dataset for machine reading comprehension, featuring real-world queries from Bing users and over 1 million passages with query-answer pairs, supporting tasks like question answering and passage ranking.
\end{itemize}
By using both FreebaseQA and MS-MARCO to fine-tune our retriever, we improve the retrieval part of our Invar-RAG more comprehensively. FreebaseQA provides complex, linguistically diverse queries matched with structured knowledge graph data, enhancing our model’s ability to retrieve fact-based answers. MS-MARCO, with its real-world user queries and passage ranking, improves performance in understanding and ranking everyday information. Together, they enhance the retriever's precision, relevance, and ability to handle a variety of natural language queries.
To fine-tune our model for the generation task, we select three other datasets (two ODQA datasets and one reading comprehension dataset).
\begin{itemize}
    \item \textbf{Wiki QA}\footnote{\url{https://huggingface.co/datasets/microsoft/wiki_qa}}: This dataset, introduced by Microsoft, consists of question and sentence pairs collected from Bing search logs. It is designed for research in open-domain question answering. The dataset contains 3,047 questions linked to Wikipedia articles, with 29,258 sentences, 1,473 of which are labeled as correct answers to the questions.
    \item \textbf{Web Question}\footnote{\url{https://huggingface.co/datasets/Stanford/web_questions}}: Released as part of research on semantic parsing, the WebQuestions dataset contains 5,810 question-answer pairs where the answers are entities from Freebase. The dataset is used to train and evaluate models for answering factual questions using structured data from Freebase.
    \item \textbf{SQuAD v2}\footnote{\url{https://huggingface.co/datasets/rajpurkar/squad_v2}}: The Stanford Question Answering Dataset (SQuAD) v2 builds upon the original SQuAD dataset by adding over 50,000 unanswerable questions that are similar to answerable ones, totaling 130,319 questions. This dataset challenges models to not only answer questions but also to determine when no answer is available in the context.
\end{itemize}
\subsection{Evaluation Datasets}
To evaluate both the retrieval and generation performance of Invar-RAG, we import three more ODQA datasets, i.e., TriviaQA, Natural Question, and PopQA, which are all large-scale public datasets that have been widely used as benchmarks in the retrieval or generation tasks~\cite{7}.
\begin{itemize}
    \item \textbf{TriviaQA}\footnote{\url{https://huggingface.co/datasets/mandarjoshi/trivia_qa}}: TriviaQA is a reading comprehension dataset that comprises more than 650,000 question-answer-evidence triples. It includes 95,000 question-answer pairs generated by trivia experts, with independent evidence documents provided—around six per question on average—offering distant supervision of high quality for answering the questions.
    \item \textbf{Natural Question}\footnote{\url{https://huggingface.co/datasets/google-research-datasets/natural_questions}}: The NQ dataset consists of questions sourced from real users, requiring QA systems to process and understand full Wikipedia articles that may or may not contain the answer. The use of actual user queries and the need to examine entire pages make this dataset more realistic and challenging compared to earlier QA datasets.
    \item \textbf{PopQA}\footnote{\url{https://huggingface.co/datasets/akariasai/PopQA}}: PopQA is a large-scale, open-domain QA dataset with 14,000 entity-centric question-answer pairs. The questions are generated by applying templates to knowledge tuples retrieved from Wikidata.
\end{itemize}
\section{Appendix B: Evaluation}
In this section, we will detail the information of our evaluation metrics and the implementation of our evaluation process.
For retrieval, to provide a comprehensive view of performance, we leverage Accuracy both truncated at five and twenty, to assess the proportion of questions where the correct answers appear in the top 5 or top 20 retrieval results, respectively. For generations, we choose the Exact Match~\cite{6} to evaluate the difference between our prediction and the ground truth. Moreover, in our evaluation process, we adopt 8-shot examples which are randomly selected from the chosen datasets following previous work~\cite{1}.
\section{Appendix C: Hyperparameter Illustration}
In this section, we will present the prime hyperparameter we used and the values we set in our fine-tuning and inferencing progress.
Tab.\ref{tab:hyper} shows the hyperparameter for fine-tuning our retrieval stage and generation stage.
\begin{table}[H]
\centering
\caption{Hyperparameter for Fine-tuning (Retrieval and Generation)}
\resizebox{\linewidth}{!}{
\begin{tabular}{ccccccc}
\toprule[1pt]
Stage &  lr\_scheduler&batch size & seq len & LoRA\_rank &LoRA\_alpha &LoRA\_dropout\\ 
\midrule
Retrieval Stage &cosine& 64 & 4096 & 16 &32& 0.05 \\
\midrule
Generation Stage &cosine& 64 & 4096 & 16 &32& 0.05 \\
\bottomrule[1pt]
\end{tabular}
}
\label{tab:hyper}
\end{table}
\section{Appendix D: Baselines Description}
(a) \textbf{BM25}~\cite{34}: A classical, sparse information retrieval method based on term frequency and inverse document frequency (TF-IDF). It ranks documents based on the occurrence of query terms, emphasizing exact term matches, making it effective in domains with well-structured data but limited in understanding semantic similarities.
(b) \textbf{BGE}~\cite{33}: A family of pre-trained embedding models designed for general Chinese text embedding. It achieves superior performance across diverse tasks by using massive datasets and advanced training techniques like contrastive learning, supporting retrieval, ranking, and classification.
(c) \textbf{Contriever}~\cite{35}: Contriever is an unsupervised dense information retriever based on contrastive learning. It is designed to overcome limitations in traditional sparse retrieval methods like BM25, particularly in zero-shot and multilingual settings. Contriever demonstrates state-of-the-art performance on multiple retrieval benchmarks, such as BEIR, and excels in cross-lingual retrieval. Its architecture relies on a bi-encoder model, which encodes queries and documents independently, and uses unsupervised contrastive learning to optimize retrieval, even without labeled data.
(d) \textbf{LLM-embedder}~\cite{36}: A unified embedding model optimized for retrieval augmentation in large language models (LLMs). It leverages multi-task fine-tuning and LLM feedback to retrieve relevant knowledge, tools, and memory, enhancing LLMs’ capabilities in knowledge-intensive tasks.
(e) \textbf{RepLLaMA}~\cite{13}: RepLLaMA is a dense retriever fine-tuned from the LLaMA-2 language model, part of a multi-stage retrieval pipeline that includes rankLLaMA as a reranker. RepLLaMA improves the effectiveness of retrieval tasks, particularly in passage and document retrieval. It outperforms smaller models, showcasing strong generalization and effectiveness in both in-domain and zero-shot settings. The model leverages both hard negatives and in-batch negatives during training.

\section{Appendix E: Ablation Study}
In this section, we will present the ablation study by gradually removing the significant components of our architecture. Specifically, we follow the variant settings in the main body and perform our ablation study on the other two evaluation ODQA datasets (Natural Question~\footnote{\url{https://huggingface.co/datasets/google-research-datasets/natural_questions}} and PopQA~\footnote{\url{https://huggingface.co/datasets/akariasai/PopQA}}), presented in Tab.\ref{tab:ab}.
\begin{table}[]
\centering
\caption{Ablation Study on Natural Question and PopQA.}
\label{tab:ab}
\footnotesize
\resizebox{\linewidth}{!}{%
\begin{tabular}{lcccccc}
\toprule[1pt]
\multirow{3}{*}{Model Variants} & \multicolumn{3}{c|}{Nature Question}                                   & \multicolumn{3}{c}{PopQA}                         \\ \cline{2-7} 
                                & \multicolumn{2}{c|}{Retrieval}      & \multicolumn{1}{c|}{Generation}  & \multicolumn{2}{c|}{Retrieval}      & Generation  \\
                                & Acc@5 & \multicolumn{1}{c|}{Acc@20} & \multicolumn{1}{c|}{Exact Match} & Acc@5 & \multicolumn{1}{c|}{Acc@20} & Exact Match \\ \hline
Default                     & 80.5 & 88.0 & 56.2 & 73.5 & 82.5 & 53.6 \\
w/o representation learning & 69.0 & 80.5 & 54.7 & 59.5 & 70.0 & 51.7 \\
w/o invariance loss         & 77.5 & 86.0 & 55.4 & 69.5 & 81.0 & 52.4 \\
w/o generative fine-tuning  & /    & /    & 53.9 & /    & /    & 50.7 \\ \bottomrule[1pt]
\end{tabular}}
\end{table}
From the experimental results, we can see that the performance of our Invar-RAG, no matter for the retrieval and the generation, shows a similar tendency to the one on TriviaQA, showing the effectiveness of our designed two-stage fine-tuning method, featuring the representation learning, invariance loss and the generation fine-tuning in one single LLM.

\bibliographystyle{ACM-Reference-Format}

\end{document}